\documentclass[pdflatex,sn-aps]{sn-jnl}% American Physical Society (APS) Reference Style
\usepackage{graphicx}%
\usepackage{multirow}%
\usepackage{amsmath,amssymb,amsfonts}%
\usepackage{amsthm}%
\usepackage{mathrsfs}%
\usepackage[title]{appendix}%
\usepackage{xcolor}%
\usepackage{textcomp}%
\usepackage{manyfoot}%
\usepackage{booktabs}%
\usepackage{algorithm}%
\usepackage{algorithmicx}%
\usepackage{algpseudocode}%
\usepackage{listings}%
\theoremstyle{thmstyleone}%
\theoremstyle{thmstyletwo}%

\theoremstyle{thmstylethree}%

\begin{document}

\title[Article Title]{Automated Journalism}

\author*[1,2]{\fnm{Wang Ngai} \sur{Yeung}}\email{justin.yeung@oii.ox.ac.uk}

\author[3,4]{\fnm{Tomás} \sur{Dodds}}\email{t.dodds.rojas@hum.leidenuniv.nl}

\affil[1]{\orgdiv{Oxford Internet Institute}, \orgname{University of Oxford}, \orgaddress{\street{1 St Giles'}, \city{Oxford}, \postcode{OX1 3JS}, \country{United Kingdom}}}

\affil[2]{\orgdiv{Leiden HumAN Lab}, \orgname{Leiden University}, \orgaddress{\street{Doelensteeg 16}, \city{Leiden}, \postcode{2311 VL}, \country{the Netherlands}}}

\affil[3]{\orgdiv{Leiden University Centre for Linguistics}, \orgname{Leiden University}, \orgaddress{\street{Reuvenplaats 3-4}, \city{Leiden}, \postcode{2311 BE}, \country{the Netherlands}}}

\affil[4]{\orgdiv{Berkman Klein Center for Internet \& Society}, \orgname{Harvard University}, \orgaddress{\street{1557 Massachusetts Ave}, \city{Cambridge}, \postcode{MA 02138}, \state{MA}, \country{United States}}}


\abstract{Developed as a response to the increasing popularity of data-driven journalism, automated journalism refers to the process of automating the collection, production, and distribution of news content and other data with the assistance of computer programs. Although the algorithmic technologies associated with automated journalism remain in the initial stage of development, early adopters have already praised the usefulness of automated journalism for generating routine news based on clean, structured data. Most noticeably, the Associated Press and The New York Times have been automating news content to cover financial and sports issues for over a decade. Nevertheless, research on automated journalism is also alerting to the dangers of using algorithms for news creation and distribution, including the possible bias behind AI systems or the human bias of those who develop computer programs. The popularization of automated news content also has important implications for the infrastructure of the newsroom, the role performance of journalists and other non-journalistic professionals, and the distribution of news content to a datafied audience.}
%%================================%%
%% Sample for structured abstract %%
%%================================%%




\maketitle

\section{Introduction}\label{sec1}

Algorithmic journalism is a container term that describes the practice of journalism that adopts techniques that involve the datafication, quantification, computationalization, and automation of several steps throughout the newsmaking process. Sometimes also termed robot journalism \cite{vanDalen2012} and automated journalism \cite{graefe2016}, algorithmic journalism attempts to automate the collection, production, and distribution of news. (For a critical account of the terminologies, see \cite{Mooshammer2022}). This entry aims to offer a brief historical overview of algorithmic journalism since the 1990s. Understanding algorithms and media technologies generally, as sociotechnical constructs, we present a range of possibilities and limitations of algorithmic journalism in contemporary newsrooms in terms of data collection, production, and distribution of news content.  

The affordances of Internet communication, such as hypertextuality and interactivity \cite{Heinonen1999}, opened new doors for journalists and general users to find information more efficiently at a meager cost. Therefore, it is unsurprising that information aggregators such as the Really Simple Syndication 1.0 (RSS) were released in 2000 when the Internet reached over 300 million users worldwide. RSS is a type of web feed that provides website updates in a standardized and computer-readable format, available for both users and applications to access. Information aggregators were also essential for journalists, as they reduced the time it took for journalists to monitor essential websites of interest. Nonetheless, information aggregators proved insufficient in the age of big data and social media platforms, when an average of 328.77 million terabytes of data are uploaded to the Internet daily.  

Indeed, the overflow of now readily available data demands an excessive amount of human labor to process and filter useful data points, especially considering that most of the content online is duplicated, unverified, or simply irrelevant. In this context, data mining techniques infused with Natural Language Processing (NLP) can prove more advantageous for media workers. The Global Database of Events, Language, and Tone or GDELT Project, created by Kalev Leetaru and Philip Schrodt, and others, is a prime example of these new emerging technologies. GDELT helps journalists monitor news media around the globe by gathering recordings from various organizations and countries while also detecting the emotions that underlie the events that have been collected. However, while GDELT is an example of an off-the-shelf tool, wealthier newsrooms have started to create their own in-house activity monitoring tools, like Reuters’ Tracer \cite{Liu2017}. 

\section{Natural Language Generation and Hyper-personalization }\label{sec2}

Despite the digitalization of journalism entering the newsrooms as early as the 1980s, journalism was not ‘algorithmic’ until the late 1990s, when a turn towards automation became manifest (See \cite{DiazNoci2013} for a historiography on this issue). For example, technologies such as the Automatic Spoken Document Retrieval, a speech-to-text algorithm conceived in the late 1980s, were leveraged to automate transcription processes of broadcasted content. Technologies for automation continue to develop inside newsrooms, and by the mid to late 2000s, full-text generation emerged as a rising tool in several media organizations. One of the most popular examples is Automated Insights, which the Associated Press later adopted as one of their main content generation tools. Other examples include the AI-generated virtual weather reporter developed by Radio Television Hong Kong (RTHK), called Aida, which started its first reporting in June 2023. 

In previous decades, the commercialization of news and rising competition between outlets prompted editors and managers to tailor content to audience segments to garner readership \cite{Dodds2023}. However, algorithmic journalism does not only involve automating the production of news content but also its distribution and marketing processes. Owing to the abundance of both implicit and explicit behavioral data collected by the news sites or offered by social media platforms themselves (e.g., Facebook, Twitter, etc.) and external data brokers (e.g., Chartbeat on The New York Times, Permutive on BBC, etc.), tailoring news content to news readers is now not only possible but encouraged by third-party stakeholders and newsrooms managers alike.  

The underlying mechanism of news tailoring in automated journalism is essentially the quantification of users’ needs and wants in the form of audience metrics. Progressively, News Recommender Systems (NRCs) that utilize user data for recommending content and maximizing click-through rate (CTR) have become ubiquitous. Google News and Apple News are the most common examples of these technologies. However, automated journalism can take personalization even further (i.e., hyper-personalization). For example, a GNI-funded project developed by the South China Morning Post attempts to cluster users based on their demographic, psychographic, and behavioral data. Setting aside ethical concerns, this project is an archetype of algorithmic journalism wherein users’ data are continuously collected, inferred, and analyzed to enhance the reader experience.  

\section{The current state of the technologies and their limitations}\label{sec3}

With the rise in popularity of Natural Language Generation (NLG), Large Language Models (LLMs), and generative AI, newsrooms are beginning to see the advantages of utilizing these technologies to automate written content. Today, the automation of the creation and curation of shorter, repetitive, and structured journalistic pieces, for instance, financial, sports, and crime news, have been rather successful, and the process of crafting such content has become considerably more streamlined (Galily, 2018). Despite the hesitation of relying on generative AI for longer, analytical pieces, advanced NLG enables a smoother and less time-consuming workflow in generative Search Engine Optimisation (SEO)-friendly content using automatic tag generation, image labeling, and alt text creation \cite{Serdouk2022}

Particularly when it comes to News Recommender System (NRS), a recent review by \cite{Raza2021} summarized current approaches and methods to recommend items to readers, including factorization models (e.g., Matrix Factorization, Bayesian Personalized Ranking, and so on) and deep learning-based systems (e.g., Neural Collaborative Filtering). Novel solutions to challenges, such as ranking newsworthiness based on timeliness, can be tackled by graph-based algorithms (e.g., sequential Browse-Graphs) and time-decay models (e.g., short-term preference model). 

However, these technologies also have limitations. For instance, although audiovisual-based generative AI is now on the rise (e.g., StableDiffusion or WaveNet), its current adoption in newsrooms has yet to be fully embraced by news organizations worldwide, which can be complicated for ethical and technical reasons. Technically, producing a photorealistic video requires considerable time to fine-tune and render compared to an AI-generated image and text. Regarding ethics, AI-generated audiovisual content inherently challenges the notion of mechanical objectivity — the reasoning of “seeing is believing” \cite{Carlson2019}. Some examples have already caused serious public debates. In early 2023, the German magazine Die Aktuelle was criticized for fabricating an interview with former F1 racer Michael Schumacher with generative AI. While this case can be easily classified as an example of deliberate misinformation, the line between a blatant lie and an AI-generated visual aid is vague.  

Another limitation of contemporary machine learning models for content generation is the differential performance across languages. A cross-language evaluative study of several generative models, such as gpt-3.5-turbo and gpt-4-32k by \cite{ahuja2023}, has elucidated that the performance of these models is best when it is prompted using high-resource languages and languages in Latin scripts. This study further showed that, for example, even if Indo-Aryan texts were translated into English before performing tasks, the performance was still comparatively worse than Germanic, Greek, and Romance language families. Such performance bias is both natural and artificial, in the sense that popular languages do have more resources to be trained on and evaluated against and are simultaneously paid less attention to simply because they are less popular. Due to such divergence, it is conceivable that newsrooms in certain regions can utilize generative AI more effectively than their counterparts. 

\section{Automated journalism impact on newsrooms’ infrastructures and changing workflows }\label{sec4}

Automated journalism has had a profound impact on the infrastructures and workflows of newsrooms, particularly their physical infrastructure. Traditionally, newsrooms relied on local data storage, utilizing computers or physical storage devices. However, with the advent of automated journalism, there has been a shift towards outsourcing data storage to cloud data warehouses offered by large platforms. Notable examples of these cloud data warehouses include Amazon Redshift and Google BigQuery. 

This transition has brought about several notable changes. Firstly, newsrooms no longer bear the responsibility of maintaining and managing their hardware for data storage. Instead, they can rely on the robust infrastructure provided by these cloud platforms, which promise scalable and secure storage solutions. This could allow newsrooms to focus more on their core journalistic activities rather than investing significant resources in maintaining physical storage systems. 

Moreover, the emergence of automated journalism has also prompted a shift in the hardware required for data analytics and model training within newsrooms. Previously, data analytics and model training were relatively rare, if not non-existent, in traditional newsroom settings. However, with the outsourcing of storage to cloud data warehouses, newsrooms now can leverage the computational power and analytical capabilities offered by these platforms. By utilizing services such as Amazon Redshift and Google BigQuery, newsrooms can perform sophisticated data analytics and train machine learning models, enabling them to uncover valuable insights and enhance their journalistic output. 

However, the transition from locally stored, processed, and distributed to outsourcing these procedures is not without criticism. Underlying these large-scale engineering systems are fragments of digital dominance \cite{MooreTambini2018} in the platform ecosystem. Indeed, newsrooms are now enabled to ‘play around’ with big data, but only possible when it is enabled by big tech companies such as Amazon and Google. The convergence of journalism and platforms has remodeled the relationship between news workers and these tech conglomerates. Not only have the platforms exerted more control over what to produce (i.e., gatekeeping), but they also have played a huge role in altering in what forms news should be produced and shared \cite{Heinrich2012}. More importantly, at the core of these platforms, in the sense of news distribution, is the quantification of users and performance based on engagement metrics. Concerning the algorithmic turn of journalism, \cite{Moyo2019} showed that although analytics-driven journalism may bring new opportunities for African newsrooms, looming was the feeling of the loss of control over the degree of whether their journalistic work actually addresses public concerns.  

The impact of automated journalism on newsrooms extends beyond physical infrastructure and encompasses changes in organizational infrastructure and workflows. One significant aspect is the rise of data journalism and the establishment of in-house data science teams. In the past, journalists did not commonly undertake quantitative and statistical data analysis. However, with the advent of automated journalism, newsrooms have recognized the value of data-driven storytelling and have formed dedicated teams to handle data journalism initiatives. Prominent examples include The Guardian US visuals team and The SFChronicle Data Team. These teams possess the expertise to explore, analyze, and visualize data, allowing for the creation of compelling and informative stories based on data-driven insights. This shift in organizational infrastructure reflects a growing recognition of the importance of data in journalism. It highlights how automation has prompted newsrooms to adapt their internal structures to accommodate these new demands. 

\section{Conclusion}\label{sec5}

In conclusion, the impact of automated journalism on newsrooms' professional infrastructures is evident in the shift from emphasizing traditional journalism skills to data-oriented technical competencies. Journalists are now expected to possess programming skills, software knowledge, and a basic understanding of machine learning models to effectively leverage automation technologies for tasks discussed in the previous section. This transformation necessitates ongoing training and professional development initiatives within newsrooms to equip journalists with the necessary skills to thrive in automated journalism. While opportunities and challenges await journalists in the age of automated journalism, more empirical research is called for to better understand how journalists can be better equipped to tackle the problem of AI unintelligibility \cite{Jones2022} and how one can reconfigure the asymmetric relationship between journalists and platforms.

\backmatter

% \bmhead{Supplementary information}

% If your article has accompanying supplementary file/s please state so here. 

% Authors reporting data from electrophoretic gels and blots should supply the full unprocessed scans for key as part of their Supplementary information. This may be requested by the editorial team/s if it is missing.

% Please refer to Journal-level guidance for any specific requirements.

\bmhead{Acknowledgements}

We thank our editors Alessandro Nai, Max Grömping and Dominique Wirz for their suggestions that have improved the quality of this encyclopedia entry. 

\bmhead{Note}

This is a preprint of an entry in Nai, A., Grömping, M., \& Wirz, D. (Eds). Elgar Encyclopedia of Political Communication. Edward Elgar Publishing.

% Some journals require declarations to be submitted in a standardised format. Please check the Instructions for Authors of the journal to which you are submitting to see if you need to complete this section. If yes, your manuscript must contain the following sections under the heading `Declarations':

% \begin{itemize}
% \item Conflict of interest/Competing interests (check journal-specific guidelines for which heading to use)
% \item Ethics approval and consent to participate
% \item Consent for publication
% \item Materials availability
% \end{itemize}

% \noindent
% If any of the sections are not relevant to your manuscript, please include the heading and write `Not applicable' for that section. 

\bigskip

\bibliography{sn-bibliography}% common bib file

\end{document}